\documentclass[aps,english,prb,twocolumn,showpacs,superscriptaddress]{revtex4}
\usepackage{graphicx}

\begin{document}

\title{Self healing of vacancy defects in single layer graphene and silicene}

\author{V. Ongun \"{O}z\c{c}elik}
\affiliation{UNAM-National Nanotechnology Research Center, Bilkent University, 06800 Ankara, Turkey}
\affiliation{Institute of Materials Science and Nanotechnology, Bilkent University, Ankara 06800, Turkey}
\author{H. Hakan Gurel}
\affiliation{UNAM-National Nanotechnology Research Center, Bilkent University, 06800 Ankara, Turkey}
\affiliation{Institute of Materials Science and Nanotechnology, Bilkent University, Ankara 06800, Turkey}
\author{S. Ciraci}
\affiliation{UNAM-National Nanotechnology Research Center, Bilkent University, 06800 Ankara, Turkey}
\affiliation{Institute of Materials Science and Nanotechnology, Bilkent University, Ankara 06800, Turkey}
\affiliation{Department of Physics, Bilkent University, Ankara 06800, Turkey}

\begin{abstract}

Self healing mechanisms of vacancy defects in graphene and silicene are studied using first principles calculations. We investigated host adatom adsorption, diffusion, vacancy formation and revealed atomistic mechanisms in the healing of single, double and triple vacancies of single layer graphene and silicene. Silicon adatom, which is adsorbed to silicene at the top site forms a dumbbell like structure by pushing one Si atom underneath. The asymmetric reconstruction of the single vacancy in graphene is induced by the magnetization through the rebonding of two dangling bonds and acquiring a significant magnetic moment through remaining unsaturated dangling bond. In silicene, three two-fold coordinated atoms surrounding the single vacancy become four-fold coordinated and nonmagnetic through rebonding. The energy gained through new bond formation becomes the driving force for the reconstruction. Under the external supply of host atoms, while the vacancy defects of graphene heal perfectly, Stone-Wales defect can form in the course of healing of silicene vacancy. The electronic and magnetic properties of suspended, single layer graphene and silicene are modified by reconstructed vacancy defects.

\end{abstract}

 \maketitle

\section{Introduction}
High mechanical strength, chemical stability, unique electronic and magnetic properties have made graphene a material of interest in diverse fields ranging from biotechnology to electronics. The honeycomb network made by planar and three-folded $sp^2$ hybrid orbitals acquires planar stability through $\pi$-$\pi$ orbital interaction and achieves high in-plane stiffness. Dirac cones provided by the linearly crossing $\pi$ and $\pi^*$ bands underlie various exceptional properties, such as ambipolar effect, massless Dirac fermion behavior etc.\cite{novoselov2004,geim2007} Silicene, a single layer, buckled honeycomb structure of Si atom, has also been demonstrated to be stable and it exhibits perfect electron-hole symmetry\cite{seymur2009, seymur2010} near the Fermi level. The structural stability is acquired by puckering of Si atoms through dehybridization of $sp^2$ orbitals. These theoretical predictions has also been confirmed by the synthesis of single layer silicene on Ag substrate.\cite{lay2010, lay2012}

Recently it was observed that, when placed in a reservoir containing enough host external atoms, graphene is able to recover and heal its vacancy defects.\cite{zan2012, robertson2013} This self healing can be perfect or result in reconstructions of the holes in different ring type structures, such as the Stone-Wales (SW) type of defect.\cite{stone86, wales98}  Vacancies in monolayer structures have attracted various experimental and theoretical studies.\cite{yazyev2010,hashimoto2004,krash2007,crespi1996,kim2011,singh2009,faccio2010}  However, the recent transmission electron microscopy (TEM) observations\cite{zan2012, robertson2013} of self healing is crucial for future device applications of graphene. Therefore, understanding the mechanisms behind this self healing process at the atomistic level and testing these mechanisms for other materials which are analogues of graphene such as graphynes,\cite{graphyne} boron nitride\cite{bn} or silicene\cite{seymur2009}, will guide future studies in the field.

Although previous theoretical models show that both graphene and silicene have perfect honeycomb structures, defects can always exist at finite temperatures. Vacancies are the most frequently observed  type of defects in crystals and they affect the mechanical and electronic properties of the materials significantly.\cite{neto2009, reina2008, banhart2011} Low defect concentration is indigenous to honeycomb structure. In two dimensional structures, vacancies are found mostly during epitaxial growth,\cite{growth} at grain boundaries,\cite{huang2011} step edges\cite{chen2010, gao2011} or any other place after a heat treatment. Owing to high formation energies of 2D honeycomb structures, substitution impurities and vacancy defects do not occur as easily as they do in other materials. On the other hand, in order to attain new functionalities, such as enhancing the catalytic activities of materials,\cite{kostov2005} vacancy defects or meshes of large holes can be created on purpose through external agents.\cite{topsakal2008,bai,balog}

\begin{figure}
\includegraphics[width=8cm]{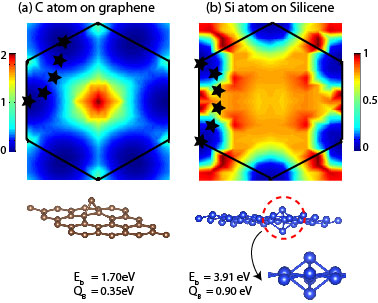}
\caption{(Color online) Energy landscapes, binding energies, $E_b$ and minimum energy barrier for diffusion, $Q_B$. (a) Energy variation and the most favorable bonding configuration of a single carbon adatom on the single layer pristine graphene. (b) Same as (a) for the Si adatom on the silicene layer.  Note that, while the graphene layer preserves its planar shape, the silicene layer and the Si adatom together reconstructs to a dumbbell structure at the bonding site. The calculations are performed using the $(8 \times 8)$ supercells. In the color code, the blue regions are energetically more favorable sites as compared to the red regions. The energy of the most favorable site is set to zero.}
\label{fig1}
\end{figure}

\begin{figure*}
\includegraphics[width=15cm]{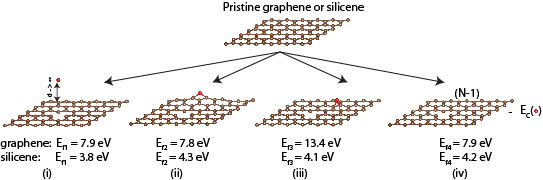}
\caption{(Color online) Schematic description for the different approaches used in the calculation of the formation energy of a single vacancy, $E_f$. The red ball corresponds to the host adatom. The vacancy formation energy for graphene was also calculated as 7.8eV,\cite{singh2009} 7.5eV\cite{faccio2010, banhart2011} and $7.0 \pm 0.5$eV\cite{thrower1978} in previous studies.}
\label{fig2}
\end{figure*}

In this study, motivated by these recent experimental evidences of graphene's ability to  self heal its vacancies, we study the formation and healing mechanisms of single, double and triple vacancy defects in free standing graphene and silicene using ab-initio calculations within density functional theory. We start with the adsorption and diffusion of single carbon and silicon adatoms on free standing pristine graphene and silicene layers, respectively. Next, we present a comparative study of the formation energy of a single vacancy based on different approaches of calculations. We show how the defected sites create a center of attraction for the adatoms in the honeycomb structure by decreasing the energy barrier of diffusion around the defects. We perform conjugate gradient calculations and finite temperature, ab-initio molecular dynamics (MD) simulations to reveal the mechanisms at the atomistic scale for both self healing, (that is healing without external atom supply), and healing via external atom supplies. We found that the energy gained by the rebonding of dangling bonds is the driving force for self-healing. The SW type of reconstructions forming during the healing process are investigated and the healing barriers of SW defects forming in graphene and silicene are compared. We finally present the effects of the vacancy defects on the electronic and magnetic structures of graphene and silicene.

\begin{figure}
\includegraphics[width=8cm]{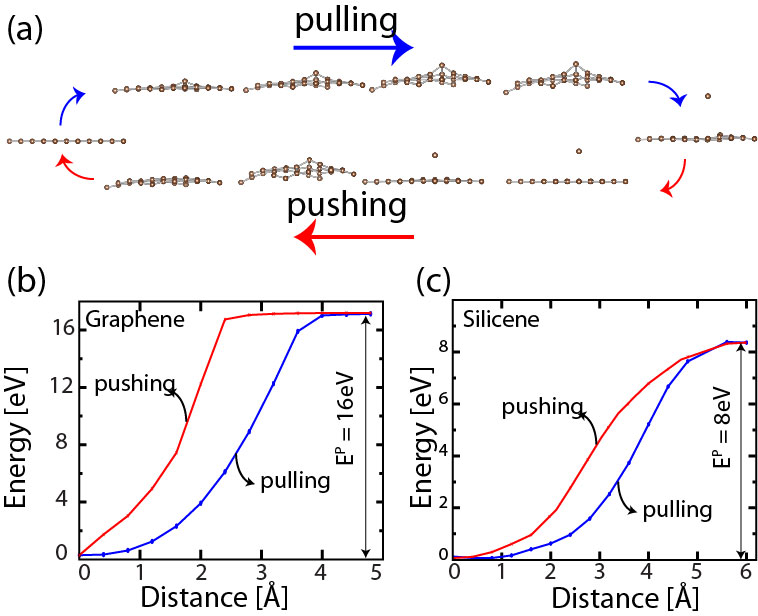}
\caption{(Color online) (a) Atomic configurations corresponding the forcibly pulling of a carbon(silicon) atom out of the (8 $\times$ 8) graphene (silicene) supercell to create a single vacancy and the reversed path obtained by pushing the same atom from a distant height towards the vacancy. (b) Variation of the total energy as a carbon atom is pulled out of graphene and pushed towards the single vacancy in graphene. (c) Same as (b) for silicene.}
\label{fig3}
\end{figure}

\section{Method}
We have performed spin polarized density functional theory calculations within generalized gradient approximation(GGA) including van der Waals corrections.\cite{grimme06} We used projector-augmented wave potentials,\cite{blochl94} and the exchange-correlation potential is approximated with Perdew-Burke-Ernzerhof (PBE) functional.\cite{pbe} A vacancy defect in the large area of honeycomb structure is represented by using the supercell method, whereby the single defect in a $(8 \times 8)$ supercell repeats itself periodically. The size of this supercell is tested to be sufficiently large to hinder defect-defect coupling. The Brillouin zone was sampled by 9x9x1 \textbf{k}-points in the Monkhorst-Pack scheme where the convergence in energy as a function of the number of \textbf{k}-points was tested. A plane-wave basis set with energy cutoff value of 550 eV was used. Atomic positions were optimized using the conjugate gradient method, where the total energy and atomic forces were minimized. The energy convergence value between two consecutive steps was chosen as $10^{-5}$ eV. A maximum force of 0.001 eV/\AA~ was allowed on each atom. Additionally,  ab initio, finite temperature MD calculations were performed where the time step was taken as 2.5 fs and the atomic velocities were renormalized to the temperature set at T = 300 K at every 40 time steps. Numerical calculations were carried out using the VASP software.\cite{vasp}

\begin{figure}
\includegraphics[width=8cm]{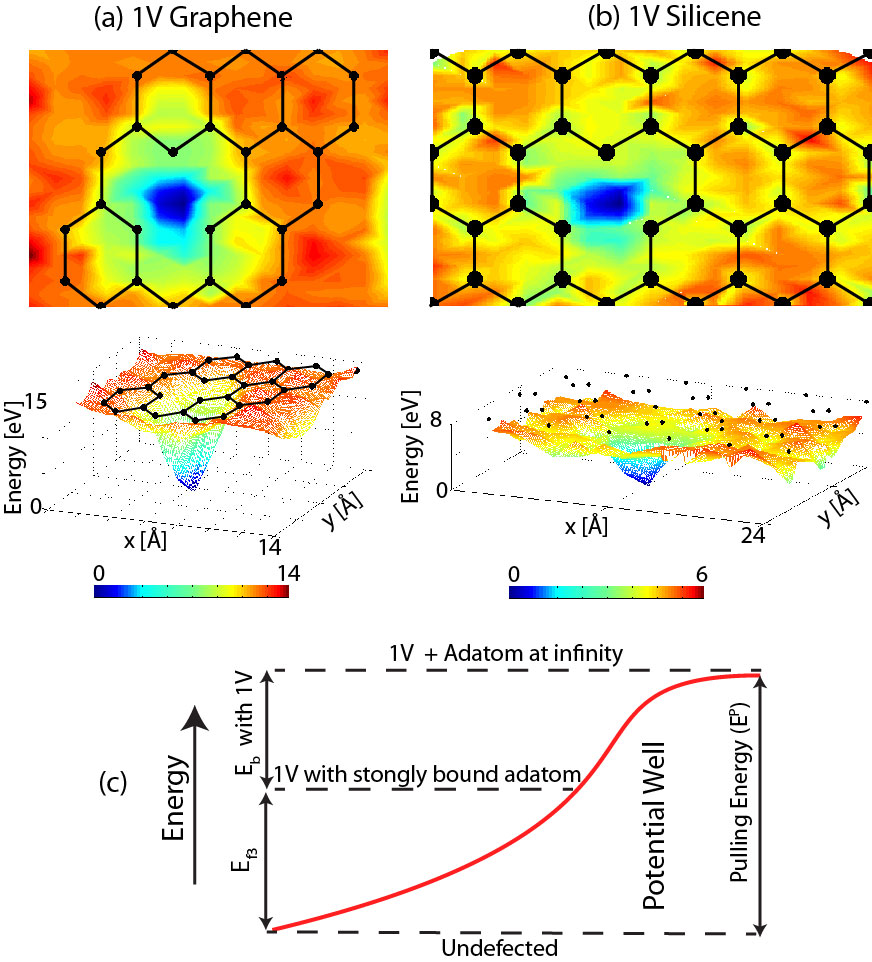}
\caption{(Color online) Energy landscapes. (a) Energy variation of a single C adatom on single layer graphene adsorbed on single layer graphene having single vacancy. The migration barrier drops significantly in the neighborhood of the defected site due to the attractive potential created at the vacancy site. (b) same of (a) for silicon adatom on the single layer silicene having a single vacancy. (c) Schematic description of the energy levels for an adatom placed at various positions on the 1V defected structure. The difference between the strong binding and the undefected structure corresponds to $E_{f3}$ value presented in Fig.~\ref{fig2}(iii).}
\label{fig4}
\end{figure}

\section{Binding and diffusion of host adatoms}
The adsorption and migration of single C(Si) adatoms on graphene(silicene) are essential for the healing of defects via ad-atoms. For this purpose we first performed spin polarized calculations on graphene and silicene supercells, which are constructed from their fully optimized primitive unit cell. All of the atoms in the structures were relaxed in all directions except for one corner atom of the supercell. The corner atom was fixed in all directions in order to prevent the substrate from slipping. The most favorable binding sites of the adatoms were determined by placing the ad-atom initially to various positions at a height of ~2\AA~ from the substrate layer and running fully self-consistent geometry optimization calculations. The adatom is kept fixed at a particular position, ($x,y)$, whereas its vertical $z$ coordinate, namely its heights from the plane, is relaxed. We repeated this calculation for a total of 500 points on the $(x,y)$-plane in one hexagon and present the final energy landscapes of carbon adatom on graphene and silicon adatom on silicene in Fig.~\ref{fig1} (a) and (b), respectively. Single carbon adatom prefers to bind to graphene exactly on the bridge site with a binding energy\cite{ebinding} of $E_b$=1.7 eV. The present value of binding energy using GGA+vdW appears to be intermediate between overbinding LDA and under-binding GGA values.\cite{binding} The C-C bond is slightly modified and the planar graphene layer is slightly bulged towards carbon adatom; nevertheless the honeycomb cage is maintained even after the adsorption of carbon atom. The most favorable binding site of the silicon adatom to silicene layer is the top site with a binding energy, $E_b$=3.9 eV. In contrast to graphene, upon binding of single Si adatom, the silicene layer modifies its original structure and is seriously distorted by the adatom. While binding to the top site, the adatom pushes the underlying Si atom of silicene layer down and a dumbbell like shape (i.e. D-structure) is formed as shown in Fig.~\ref{fig1} (b), which is the energetically most favorable geometry for a silicon atom adsorbed on silicene. We note that in our earlier work it was found that the similar dumbbell structure can occur also on graphene, even if it is not an equilibrium structure.\cite{canethem} Much recently, ($\sqrt{3}$x$\sqrt{3}$) periodic structure of dumbbell on silicene was found to be energetically more favorable than pristine silicene.\cite{kaltsas} Here, the D-structure is formed as a defect state of a Si adatom migrating on pristine silicene. The formation of D-structure is exothermic and occurs without any energy barrier once Si adatom is at the T-site.

The migration paths with minimum energy barrier, $Q_B$ of C(Si) adatom  on graphene(silicene) are marked with stars in Fig.~\ref{fig1} (a) and (b). Accordingly, the minimum energy barrier $Q_B$ for a single C adatom to migrate from one bridge site to another bridge site is 0.35eV.\cite{migration} Similarly, the corresponding minimum energy barrier is calculated as $Q_B$=0.90 eV for Si adatom on silicene.

\subsection{Formation energy of a single vacancy}
The formation energy $E_f$ of a single vacancy and the ambient temperature determines the equilibrium vacancy concentration in graphene and silicene. Various approaches have been used in the earlier works\cite{singh2009,faccio2010,gillan89} dealing with the formation energy of a single vacancy in graphene. In this section, we present a comparative study for the calculations of vacancy formation energy. Four different approaches used in the present study is illustrated in Fig.~\ref{fig2}: (i) We forcibly pull an atom out of the perfect structure such that it completely detaches from the lattice;  (ii) we remove an atom from its original position and let it migrate and bind to the edge of the flake as if the flake further grows, (iii) we remove an atom from the lattice and let it bind to the bridge or to the top site for graphene and silicene, (iv) we follow the approach used in Ref[\onlinecite{gillan89}] and calculate the formation energy using the same calculation parameters of this study.

In the first approach (i), we completely detach one C(Si) atom from a perfect $(8 \times 8)$  supercell of graphene(silicene). To do this, we systematically pull out a C(Si) atom from the graphene(silicene) layer and examine the variation of the total energy as we increase the distance of the pulled atom from the layer. Fig.~\ref{fig3} shows the variation of total energy as a function of the atom's height from the layer, where the optimized energy of the defect-free structure, $E_{T}^H$, is set to zero. As the pulled atom gets further away, the total energy increases gradually. Once the pulled atom is far away from the graphene(silicene) layer, the total energy saturates at a constant value. This shows that the pulled atom is completely detached. The energy difference between the detached configuration and the initial configuration gives us the energy of the forced defect formation by pulling and completely removing an atom, $E^P$, which we calculate as $\sim$16eV for graphene and $\sim$8eV for silicene. In fact, this value is equal to the difference between the total energy values of perfect pristine graphene(silicene), $E_T^H$, and a defected graphene(silicene), $E_{T}[1V]$,  plus one single carbon(silicene) atom, $E_{T}[A]$, namely $E_{p}=E_T[1V] + E_{T}[A] - E_{T}^{H}$. The calculated $E_p$ of graphene is about two times larger then that of silicene indicating that it is harder to form vacancy defects in graphene as compared to silicene. One can retrieve the vacancy formation energy using the cohesive energy of C(Si) in graphene(silicene), $E_C$, as $E_{{f}_1}=E_{p}-E_{C}$. Using the cohesive energy of C(Si) in graphene(silicene) as $E_C$=8.1 (4.2) eV, the first approach leads to the value $E_{{f}_1}$=7.9(3.8) eV for the vacancy formation energy for graphene(silicene).

We also repeat the same analysis by pushing this detached atom back into its place. Although the energy path is different, the change in energy is the same as shown in Fig.~\ref{fig3}(b) and (c). The hysteresis, namely the difference in the pulling and pushing curves is due to the difference in the strain induced bulging of the graphene layer in the course pulling out and pushing in of the host atom. The bulging is initially not present as we push an adatom into a vacant site, until the pushed atom starts to interact with the graphene or silicene layer.

In the second approach (ii), we follow the formal definition of the vacancy formation energy; namely as the energy required to remove one atom from the crystal and add to the surface.\cite{kittel}. This approach is more suitable for the vacancy defect in finite graphene(silicene) flakes.  To this end we create a vacancy in graphene (silicene) by removing a single carbon (Si) atom from an $(8\times8)$ flake of graphene and subsequently relax the defected flake. Thereafter, we let the removed atom to attach to the edge of the flake so that its size is extended. Accordingly, the formation energy of a finite flake is calculated by the difference $ E_{{f}_2} = E_T[1V+1A] - E_T^H $ in terms of the optimized total energy of the flake including one vacancy and one additional host atom attached at the edge and that of perfect flake, respectively. We found $E_{{f}_2}$ is 7.8 eV for graphene and 4.3 eV for silicene. We believe that, this is the most realistic and accurate way of calculating defect formation energy for a finite flake. It is in compliance with earlier theoretical studies predicting that carbon adatom on a small flake tends to migrate to the edge and to attach to the edge as if the flake is growing.\cite{migration, ongun_chain}

Alternatively, in the third approach (iii) the removed atom may also fall into a local potential well and be trapped behind a high diffusion barrier which blocks its migration to the edge. This is most likely to happen if it attaches to its most favorable binding site on the lattice far from the edges of a large flake. Therefore, as a next step rather than letting the removed atom migrate to the edge, we attach it to its most favorable site on the lattice, which is the bridge site for graphene and top site for silicene. We treat this intermediate case using $(8 \times 8)$ supercell geometry and calculate the formation energy as the energy difference given by $E_{{f}_3}= E_{T}[1V+1A_{add}] - E_T^H$ which is 13.47 eV for graphene and 4.13 eV for silicene. Here $E_{T}[1V+1A_{add}]$ is the total energy of the defected supercell where the removed host atom is bound to graphene(silicene) as an adatom (i.e. bridge bonded carbon on graphene/dumbbell structure on silicene). Here we make the following comments: The formation energy of vacancy in graphene is larger in this intermediate state since the cohesive energy of carbon atom in graphene is much larger than the bridge bonded carbon. Also the binding energy of C adatom changes due to the magnetic ground state of the underlying defected supercell, which is further discussed in section IV-A. The entropy of disorder in the formation of vacancy is different for the approach (ii) and (iii).

Finally(iv) we calculate the formation energy of vacancy as suggested in previous theoretical studies.\cite{gillan89} Accordingly, the formation energy is obtained from the expression, $E_{{f}_4}= E_{T}[1V,N-1]-E_{T}^{H} \times (N-1)/N$, where $E_{T}[1V,N-1]$ is the optimized total energy of a  $(\sqrt{\frac{N}{2}} \times \sqrt{\frac{N}{2}})$ supercell containing 1V and $N-1$ C(Si) atoms, and $E_{T}^H$ is the optimized total energy of the same supercell of pristine graphene(silicene). We calculated the formation energy of a single vacancy of graphene(silicene) for supercells $(n \times n)$ for $n=\sqrt{\frac{N}{2}}$=5,6,7,8 and 10 to be respectively, 7.96(4.15) eV, 7.98(4.16) eV, 7.95(4.16) eV, 7.94(4.18) eV and 7.95(4.18) eV. These values are in fair agreement with previous theoretical and experimental studies,\cite{singh2009,faccio2010,thrower1978,banhart2011} and also indicates that our calculations performed using $(8\times8)$ supercell is converged. Present values of formation energies differ from previous studies by 6\% due to different computational parameters, different sizes of supercells and basis sets used in their calculations. We also note that the calculation of vacancy formation energy using different approaches are in conformity.

\subsection{Migration of adatoms on defected lattice}
We next focus on how the existing vacancy defects alter the diffusion barriers of C and Si adatoms . For this purpose we performed large scale calculations of the energy landscape of adatoms on the lattice. We used $(8 \times 8)$ supercells of graphene and silicene, which contain a single vacancy defect. Similar to what we did for the perfect structure, we place a single C or Si adatom on various places of the $(6 \times 6)$ supercell. All the atoms in the graphene and silicene layers were fully relaxed except for one corner atom, in order to avoid slipping. While the adatoms are fixed laterally at the ($x,y$)-position on the graphene (silicene) layer they are fully relaxed in the $z$ direction. Instead of scanning through a single hexagonal portion, this time we scan the energy landscape over the entire supercell by placing the adatom at 525 different places. This way, we observe the effect of the vacancy on the migration of adatom.

As seen in Fig.~\ref{fig4}, at places further away from the defected site, the energy profile resembles the profile of the perfect case shown in Fig.~\ref{fig1}. However, at the close proximity of the defected site, that is when the distance of the adatom to the vacancy is less than $\sim$ 3\AA, the diffusion barrier $Q_B$ starts to decrease gradually and dips into a minimum when it is on the vacant site. Therefore, once the adatom gets within a close neighborhood of the defected site, it is attracted by the defect, which eventually enhances the healing of the vacancy defect. In this respect, the energy profiles presented in Fig.~\ref{fig4} can be viewed as potential wells which attract the adatoms for healing of defects. The depths of these wells correspond to the difference between the maximum energy points on the supercell and the minimum energy point(the defected site). Note that the maximum energy points are not the most favorable binding sites, but rather loose binding sites. Also, these energy profiles bear upon the vacancy formation energies calculated by using the approach (iii) above. If one subtracts the energy at the defect site i.e. the minimum of the well from the energy at the most favorable site (bridge site of graphene or the top site of silicene corresponding to D structure) one obtains the formation energies of vacancies $E_{{f}_3}$=13.47 eV (4.13 eV). Apparently, it is harder to create a vacancy defect in graphene, which in turn leads to a higher attractive potential at the defected site and also stronger tendency for healing. It is worth to note that although the attractive potential of a vacancy in the graphene layer is deeper than the potential of a vacancy in the silicene layer, the distance ranges over which these potentials interact with the adatoms are similar. In other words, the diffusion barriers start to decreases when the distance between the adatom and the defected site is less than $\sim$ 3~\AA~for both cases. Defect healing process of a vacant site is expected to occur if the adatom is present in this low barrier region.

\begin{figure*}
\includegraphics[width=16cm]{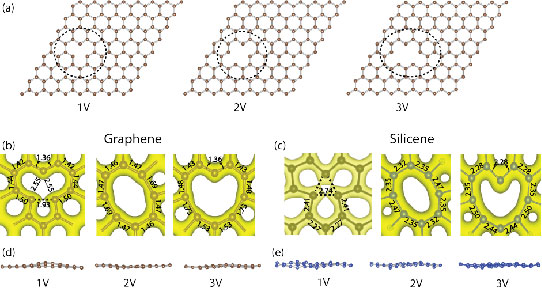}
\caption{(Color online) Self-healing and reconstruction of vacancy defects in graphene and silicene. (a) Initial unrelaxed configurations of the single(1V), double(2V) and triple(3V) vacancies.  (b-c) Optimized atomic configurations and the isosurfaces of the total charge densities of the reconstructed vacancy defects in graphene and silicene. (d-e) Molecular dynamics simulation results of defected graphene structures at 300 K. The wrinkling occurs in the MD simulations due to the effect of the temperature. The isosurface value is taken as $0.01 e/ \AA^3$.  The defects are treated in an $(8\times8)$ supercell.}

\label{fig5}
\end{figure*}

\section{Healing of vacancies in graphene and silicene}
Prime objective of the present study is to reveal the atomistic mechanisms of the healing processes observed in Ref's[\onlinecite{zan2012,robertson2013}]. In this section, we consider both self-healing/reconstruction without external atom supply and healing with external host atom supply.

\subsection{Self-healing and reconstruction of vacancy defects}
For the self-healing or reconstruction, the structures are left to heal and reorient  by themselves immediately after a vacancy defect is created in the pristine graphene and silicene. Single, double and triple vacancies are created in the perfect honeycomb structures and healing is examined without the presence of external atoms. The defected single layers are relaxed by both SCF conjugate gradient calculations and ab-initio MD calculations at 300K.

We present the unrelaxed geometry of 1V, 2V and 3V defects, as well as their final optimized geometries together with their total charges densities in Fig.~\ref{fig5}(a)-(c). The optimized structures indicate that the defected structures shrink and the bonds reorient to close the vacancies. However, this shrinking and reconstruction is not sufficient to completely heal these defected structures.

We perform both spin polarized and spin unpolarized calculations for all vacancy defects in graphene and silicene. When spin unpolarized calculations are performed, graphene and silicene with single vacancy preserves its initial symmetric geometry presented in Fig.~\ref{fig5}(a). Single vacancy in graphene has three atoms, each of them is two fold coordinated and has a $sp^2$-dangling bond. However, for graphene the magnetic ground state of this single vacancy is energetically more favorable as compared to  the nonmagnetic state by an energy difference of 0.8 eV. Upon reconstruction, which is induced by magnetization, the atoms around the single vacancy of graphene reconstructs in an asymmetric way relative to the initial configuration. Two of the three atoms around the defected site form a C-C bond, as shown by the dashed lines in Fig.~\ref{fig5}(b). This result is consistent with recent experimental observations,\cite{robertson2013} which showed that the asymmetrically reconstructed single vacancy oscillates between three favorable configurations. In other words, of the initial three dangling C bonds, any two of them are equally likely to rebind and hence to form a C-C bond.

In the case of the single vacancy in silicene, the nonmagnetic ground state is always more favorable. In the unrelaxed defected structure, the atoms surrounding the vacancy have three dangling bonds. Upon relaxation, these atoms move towards the center of the vacancy and form bonds with each other as shown by  the dashed lines in Fig.~\ref{fig5}(c). As opposed to the case in graphene, these bonds have the same length and are all stable. As a result of the reconstruction, three pentagons and three hexagon occur around the vacancy. At the end, existing dangling bonds are saturated and three two-fold coordinated Si atoms surrounding the vacancy become four-fold coordinated. The resulting structure is symmetrical around the vacancy defect. This geometrical symmetry also leads to a symmetry in the distribution of magnetic moments in the lattice, making the ground state singlet. Since the magnetic moments cancel out each other in singlet ground state, the resulting structure has zero net spin and zero magnetic moment.\cite{sakurai, fan2009, ongun_chain} This is dramatically different from the magnetic ground state of graphene with single vacancy.

\begin{figure*}
\includegraphics[width=15cm]{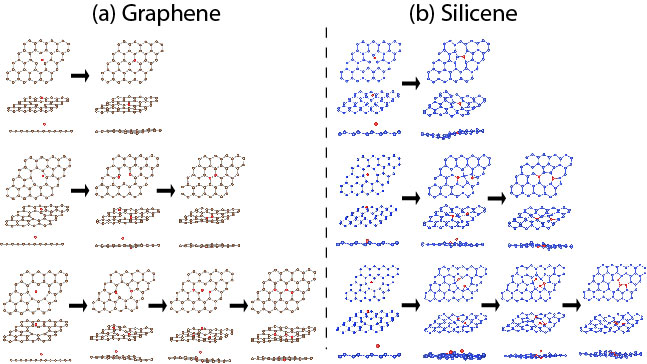}
\caption{(Color online) Molecular dynamics simulations of healing under the host adatom supply for single, double, and triple vacancies in graphene and silicene. Carbon, silicon and host adatoms are shown with brown, blue and red balls, respectively. While graphene heals its defects perfectly for each type of vacancy, defected silicene may end up with Stone-Wales(SW) type of defects in the course of healing assisted with Si adatoms.}
\label{fig6}
\end{figure*}

In the unrelaxed double vacancy, there are four two-fold atoms surrounding the hole, each of them has $sp^2$-dangling bond oozing towards the center of the hole. These surrounding atoms are reconstructed in such a way that four dangling $sp^2$-bonds are saturated in pairs to form two C-C or Si-Si bonds and hence the energy of the defected graphene or silicene is lowered due to the formation of two new bonds. This reconstruction is reminiscent of the reconstruction occurring on the $(2 \times 1)$ reconstruction on the ideal Si$(001)$ surface. Eventually, a ring of eight carbon(silicon) atoms forms in graphene(silicene), which are surrounded by six hexagons and two pentagons. The atoms around the eightfold ring are made from saturated bonds and hence are chemically rather inactive.

As for the triple vacancy, it has uneven number of two-fold coordinated atoms surrounding the hole of vacancy, each having a dangling bond. From five dangling bonds four of them are combined in pairs to form two C-C or Si-Si bonds. Since the fifth dangling bond cannot be paired, it continues to be dangling. At the end, a heart-like hole is formed upon reconstruction which is surrounded by 8 hexagons, 2 pentagons and one dangling bond oozing towards the center. This dangling $sp^2$-bond makes the healed triple vacancy chemically active.  We calculate the reconstruction energies by taking the energy difference between the reconstructed and unrelaxed structures. The results presented in Table~\ref{table1} indicate that all of these reconstructions are exothermic processes. This energy is the driving force for the healing process. The calculated magnetic moments of the supercells of these vacancy defects are also presented in Table~\ref{table1}. Note that, all structures have zero net magnetic moment except for 1V and 3V graphene. The electronic properties of these structures are further investigated in the forthcoming section.

\begin{table}
\caption{Reconstruction energies, $E_R$ (eV per defect), and magnetic moments, $\mu$ ($\mu_B$ per defect), for a single layers of graphene and silicene having a defect. Single vacancy (1V), double vacancy (2V), triple (3V) vacancy and Stone-Wales(SW) defects are treated in an $(8 \times 8)$ supercell. $E_R$ value for the Stone-Wales (SW) defect corresponds to the energy difference between the SW defected state and completely healed state consisting of regular hexagons. Magnetic moments per number of atoms are given in parenthesis.}
\label{table1}
\begin{center}
\begin{tabular}{ccccc}
\hline  \hline
& \multicolumn{2}{c}{$E_R$}  & \multicolumn{2}{c}{$\mu$} \\
Defect Type & Graphene & Silicene & Graphene & Silicene \\
\hline
1V & 0.29 & 0.46 & 1.52 (0.012) & 0 \\
2V & 2.62 & 1.83 & 0 & 0 \\
3V & 2.27 & 1.89 & 1.01 (0.008) & 0 \\
SW & 5.70 & 2.12 & 0 & 0 \\
\hline
\hline
\end{tabular}
\end{center}
\end{table}

For graphene and silicene, the predictions of SCF conjugate gradient and ab-initio MD calculations at 300 K on the atomic configurations are similar, except for the wrinkling occurring in the MD simulations due to the effect of the temperature. In Fig.~\ref{fig5} (d) and (e) we show the side views of the final configurations obtained after the MD simulations. In both SCF-CG and MD calculations, although perfect hexagons do not form, these vacancies reconstruct in the form of larger rings, as shown in Fig.~\ref{fig5}. To attain the perfect healing we next start to add host adatoms by placing them externally and randomly around the defects.

\begin{figure}
\includegraphics[width=8cm]{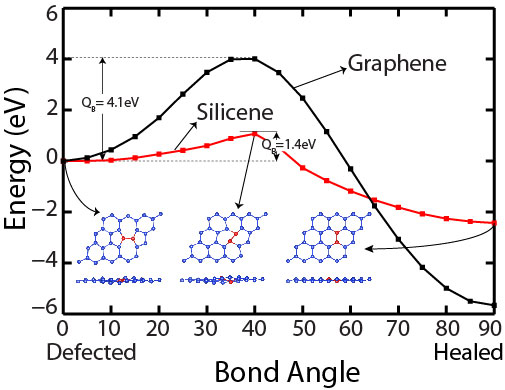}
\caption{(Color online) Energetics of the healing paths of SW defects in graphene and silicene are shown in black and red lines, respectively. The minimum energy barriers of the healing of SW defects (the energy difference between SW defected and perfect states) are 4.1 (5.7) eV and 1.4 (2.1) eV for graphene and silicene, respectively.}

\label{fig7}
\end{figure}

\subsection{Healing with external atoms}
In this section, we demonstrate that perfect healing may occur if host ad-atoms are supplied externally. To this end, as a second step, we assume that the external atoms are supplied to the system at the close proximity of the defect and investigate their effects on the healing process. We perform SCF conjugate gradient calculations. We initially let the defected structure relax without ad-atoms as explained in the previous section. Subsequently, we supply external host ad-atoms to the defected structures and let the systems relax again. As the new atoms are introduced to the system, they eventually fill the vacancies and start to heal graphene and silicene. For graphene, the vacancies always heal perfectly as long as there are sufficient number of additional host adatoms to compensate the missing atoms in the vacancy defects. For example, three additional carbon atoms are sufficient to perfectly heal a triple vacancy. Apparently, the healing of reconstructed graphene defects occurs spontaneously. The situation is slightly different in the case reconstructed silicene defects: The final structure of silicene depends on the initial positions of the adatoms supplied. In the presence of external Si adatoms, even though the single vacancy in silicene always heals spontaneously and perfectly, the double and triple vacancies may either form Stone-Wales defects or heal perfectly.

To further investigate this situation attained by conjugate gradient, we perform ab-initio MD simulations at 300 K. The MD calculations were run for 5ps before the next atom is supplied from a random position. Single atoms supplied on top of the layers eventually move towards the hole region. To by-pass the time required for the diffusion of single atom to the vacant area, the adatoms were supplied at a close proximity of the defected region at random locations. The healing processes of graphene and silicene are shown in the first and second columns of Fig.~\ref{fig6}. The adatoms move to the positions of the removed atoms and eventually heal the system. This healing is perfectly done for the case of graphene, due to the high attractive potential at the vacant site. However, the layers cannot heal perfectly in silicene; Stone-Wales type defects can occur in the final structures of the silicene layers. This, together with the results of the conjugate gradient calculations suggests that the strength of the attractive potential at the defect site is the driving force for the healing. This driving force is lower in silicene and hence the healing can terminate with SW defect.

For a perfect healing of silicene, these SW defects should heal to ordinary hexagons. The healing of a SW defect to two regular hexagons can be achieved by the rotation of the bond between the heptagon and the pentagon forming the defect. A system consisting of only hexagons is energetically more favorable than a system containing SW defects. However there exists an energy barrier which is needed to be overcame for the transformation of a SW defect to two perfect hexagons. In Fig.~\ref{fig7} we calculate this energy barrier for graphene and silicene to be 4.1 eV and 1.4 eV, respectively. On the other hand, the energy differences between SW defect and healed state consisting of regular hexagons are calculated to be 5.7 eV 2.1 eV for graphene and silicene, respectively.

\begin{figure}
\includegraphics[width=8cm]{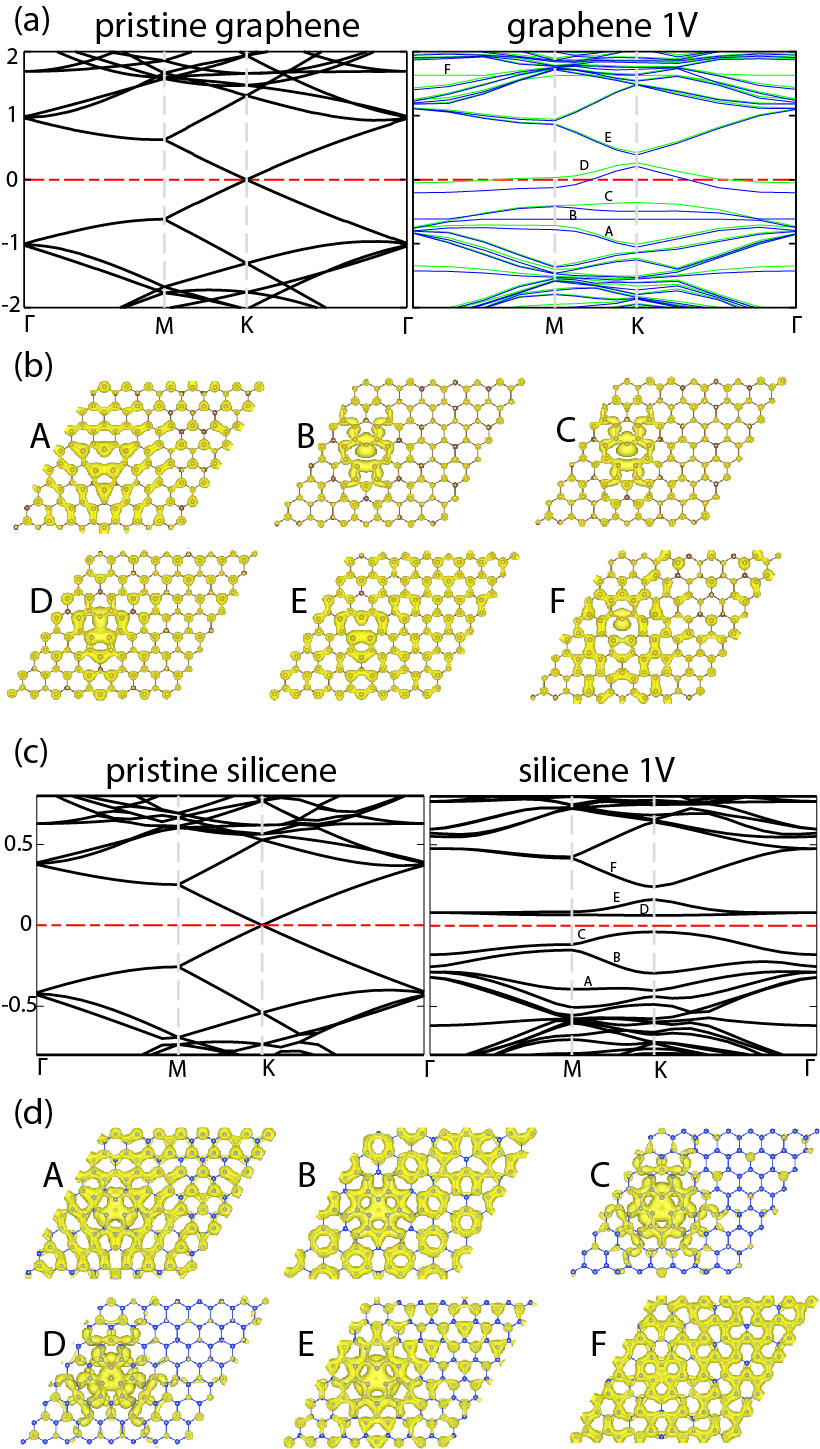}
\caption{(Color online) Energy band structure and isosurface charge densities of graphene and silicene having a single vacancy (1V) in the $(8 \times 8)$ supercell. Spin degenerate, spin up and spin down bands are shown by black, green and blue lines, respectively. The  zero of energy is set to the Fermi level. The energy bands of pristine graphene and silicene folded to the Brillouin zone of $(8 \times 8)$ supercell are presented in left panels for the sake of comparison. (a) Energy band structures of pristine and defected graphene. Owing to the magnetic ground state of the defected graphene the bands are split. (b) Integrated charge density isosurfaces of the bands A-F. Charge densities of flat bands are localized at the defect region. (c) Energy band structures of pristine and defected silicene. (d) Integrated charge density isosurfaces of the bands A-F of silicene.}

\label{fig8}
\end{figure}

\begin{figure*}
\includegraphics[width=15cm]{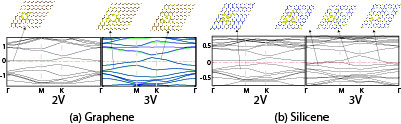}
\caption{(Color online) Energy band structure and integrated charge densities of specific bands associated with defects of graphene and silicene having a double (2V) and triple (3V) vacancy in the $(8 \times 8)$ supercell. Spin degenerate, spin up and spin down bands are shown by black, green and blue lines, respectively. The  zero of energy is set to the Fermi level. (a) Graphene including 2V and 3V. Owing to the magnetic ground state of 3V, the spin-up and spin-down bands are split. (b) Silicene including 2V and 3V.}

\label{fig9}
\end{figure*}

\section{Electronic and magnetic properties}
Vacancy defects are localized states; they give rise to localized states in the gap or resonance states in the band continua. In low defect concentrations the bands of graphene or silicene become undisturbed; the defect states occur as donor or acceptor states in the band gap or they mix with the bands of pristine graphene or silicene to change them into resonance states in the band continua. In our study we treat the vacancy defects using periodic boundary conditions, whereby the vacancy defect under study has repeated periodically in the $(8 \times 8)$ supercells. Therefore, within this framework, the concentration of single vacancy is very high and $C_{1V}$=1/128. Under these conditions the states related with the defect form energy bands like the bands of host graphene or silicene and attain dispersion depending on the size of the supercell and their coupling with defects in adjacent cells. In addition, for graphene and silicene, if the mesh of vacancy defects in the 2D hexagonal lattice breaks the specific symmetries of parent pristine graphene and silicene structures the linearly crossing bands at the Fermi level open a gap.\cite{hasan} Under these circumstances, one cannot reproduce the electronic structure of a realistic system having very low vacancy concentration by using supercell method. Of course, even if the calculated structure converges to the realistic system with increasing size of supercell, the limited computation capacity does not allow us to use very large supercells. In this study, we nevertheless aim to reveal essential aspects of the electronic structure using the present supercell method. In particular, we expect that a single layer graphene or silicene with very low vacancy concentration still have linearly crossing bands, and localized vacancy states corresponding to the flat impurity bands in the band gaps.

First we focus on the magnetic and electronic properties of the defected and reconstructed structures. We perform both spin polarized and spin unpolarized energy minimization calculations for the reconstructed single, double and triple vacancies in graphene and silicene. The electronic structure and charge density of relevant bands of graphene and silicene having 1V in the $(8 \times 8)$ supercell are presented in Fig.~\ref{fig8}. For graphene, 1V magnetic ground states, since the number of carbon atoms in two sublattices differ by one leading to unpaired $\pi$ electrons.\cite{lieb} The magnetic state is more favorable in energy by 0.47 eV. For single vacancy, we find a net magnetic moment of 0.012 $\mu_B$ per atom (or 1.52 $\mu_B$ per vacancy). The variation of magnetic moment with the size of the supercell fits the trend presented in recent studies.\cite{yazyev2007, faccio2010, singh2009} As a result, electrons spin degeneracy of the bands are broken and bands split as seen in Fig.~\ref{fig8} (a). Due to symmetry of the supercell having one single vacancy, the linearly crossing bands are split and they are raised slightly above the Fermi level. The $\pi$ and $\pi^*$ bands around the Fermi level mix with the orbitals of vacancy. The states associated with the dangling bond and reconstructed C-C bond of vacancy occurs near the top of valence band and in the conduction band and appear as flat bands indicated as B, C and F. The charge densities associated these bands are localized as shown in Fig.~\ref{fig8} (b). The overall features of bands are in agreement with the that calculated by Faccio et al.\cite{faccio2010} using local basis set.

The bands corresponding to the nonmagnetic ground state of silicene having single vacancy in the $(8 \times 8)$ supercell is presented in Fig.~\ref{fig8} (c). Because of the symmetry breaking due to vacancy, the linearly crossing bands open a band gap. The bands indicated by C and D correspond to the localized states associated with the four-fold coordinated bonds of the atoms surrounding the vacancy. In addition, the orbitals of these bonds mix with the $\pi$ and $\pi^*$ states.

Graphene with double vacancy has zero net magnetic moment since two sublattice have equal number of atoms and it doesn't have any dangling bonds in the reconstructed configuration as shown in Fig.~\ref{fig5}(b). However, the triple vacancy has one dangling bond and unpaired electrons as in the single vacancy. Thus, the unpaired electrons create a net magnetic moment in the overall structure as also indicated by Lieb's theorem.\cite{lieb}  On the other hand, as presented in Fig.~\ref{fig9} (b), the reconstructed 2V and 3V of silicene have nonmagnetic ground states due to the buckled structure of silicene as opposed to planar graphene.

\section{Conclusion}
Vacancy defects in single layer graphene and silicene exhibit unusual properties emerged from their honeycomb structures and rotation symmetries. Experimentally, it is revealed that the concentration of vacancy defects in graphene is low and the  structure has a tendency to self-heal. Using first-principles plane wave calculations within the density functional theory, we show the atomistic mechanisms behind the healing of vacancy defects occurring in graphene and silicene. We investigated adsorption and migration of host adatoms on graphene and silicene and presented a comparative study of the methods used for the calculation of vacancy formation energies in these single layer honeycomb structures. We also show that in graphene the magnetization induces reconstruction of the single vacancy, whereby two dangling bonds rebond to lower the energy of the system while the remaining dangling bond acquires magnetic moment. The rebondings of atoms surrounding the vacancy drive the reconstruction of self-healing. We showed that how these vacant sites create attraction for the host adatoms present in the medium. This attraction decreases the energy barrier for the migration of host adatoms and enables the healing of defects. Healing from a vacancy to a perfect hexagonal structure is more likely to occur as more adatom supply is present. During the healing of vacant sites, the supercell may also reconstruct to form Stone-Wales type of defect first which then recover by a bond rotation process. The energetics and the activation barrier of this mechanism is presented. The energy barrier for the bond rotation process is lower in silicene as compared to graphene by $\sim 3eV$. We believe that the atomistic mechanisms behind defect formation and healing results presented here will guide future self healing studies in graphene and other materials analogues to graphene.

\section{Acknowledgement}
The computational resources have been provided by TUBITAK ULAKBIM, High Performance and Grid Computing Center (TR-Grid e-Infrastructure) and UYBHM at Istanbul Technical University through Grant No. 2-024-2007. This work was supported partially by the Academy of Sciences of Turkey(TUBA).


\begin{thebibliography}{15}
\bibitem{ebinding}
The binding energy is calculated using the expression, $E_b=E_{T}^{H}+E_{T}[A]-E_{T}[H+A]$ in terms of the total energy of bare graphene or silicene (denoted by $H$) $E_{T}^H$, the total energy of the free atom, $E_{T}[A]$, and the total energy of an adatom $A$ adsorbed $H$, $E_{T}[H+A]$. All total energies are optimized in a $(8 \times 8)$ supercell.
\end{thebibliography}
\end{document}